\documentclass[conference]{IEEEtran}
\usepackage{packages}
\begin{document}

\title{No Two Developers Think Alike: How Problem-Solving Styles and Experience Shape Needs in Conversational Interaction with Copilot}

\author{%
\IEEEauthorblockN{Jonan Richards \orcidlink{0009-0007-1218-8599}}
\IEEEauthorblockA{%
\textit{Radboud University}\\
Nijmegen, the Netherlands
}
\and
\IEEEauthorblockN{Bruno Alves de Oliveira \orcidlink{0009-0003-2026-4361}}
\IEEEauthorblockA{%
\textit{Federal University of Technology -- Paraná}\\
Campo Mourão, Brazil
}
\and
\IEEEauthorblockN{Iury Oliveira}
\IEEEauthorblockA{%
\textit{Microsoft}\\
São Paulo, Brazil
}
\and[\hfill\mbox{}\par\mbox{}\hfill]
\IEEEauthorblockN{Igor Wiese \orcidlink{0000-0001-9943-5570}}
\IEEEauthorblockA{%
\textit{Federal University of Technology -- Paraná}\\
Campo Mourão, Brazil
}
\and
\IEEEauthorblockN{Mairieli Wessel \orcidlink{0000-0001-8619-726X}}
\IEEEauthorblockA{%
\textit{Radboud University}\\
Nijmegen, the Netherlands
}
}


\maketitle

\begin{abstract}
Conversational LLM-based ``programming assistants'' provide a range of benefits to developers. However, recent studies demonstrate the variety in individual developers' needs regarding programming assistants, and challenges encountered by only specific groups of developers. In this study, we explore the role of cognitive diversity in shaping interactions with GitHub Copilot chat. Through a mixed-methods think aloud study with 27 professional developers and students, we characterize 5 distinct ``interaction modes'' and 10 underlying needs in developers' interactions, forming a conceptual model. We characterize links between these modes, needs, and developers' problem-solving styles and experience profiles, showing how cognitive diversity may shape developers' interactions. We provide insights and recommendations for researchers and practitioners on how to design, research, and employ programming assistants to better account for diverse developer needs.
\end{abstract}


\section{Introduction}
Large Language Models (LLMs) are increasingly demonstrating strong performance across a wide range of code-centric software engineering (SE) tasks, including code generation, bug detection, and supporting developers' code understanding~\cite{hou2024LargeLanguageModels}. Conversational (chat-based) LLM tools for SE, hereafter referred to as ``programming assistants'', are especially promising. Conversational interaction supports iterative refinement and in turn productivity~\cite{austin2021ProgramSynthesisLarge}, supports multiple SE activities in a single interface~\cite{ross2023ProgrammersAssistantConversational}, and, notably, is missed by developers when absent~\cite{liang2024LargeScaleSurveyUsability}. These observations suggest that programming assistants provide not only technical capabilities but also support human aspects of software development. Examples include GitHub Copilot chat\footnote{\url{https://github.com/features/copilot}} and Tabnine\footnote{\url{https://www.tabnine.com/ai-chat/}}, as well as more general-purpose conversational assistants like ChatGPT\footnote{\url{https://openai.com/index/chatgpt/}}, which developers frequently use for programming-related tasks~\cite{xiao2024DevGPTStudyingDeveloperChatGPT}.

Cognitive diversity impacts a range processes including decision making, productivity, and collaboration~\cite{mello2015CognitiveDiversityTeams}. In software engineering, cognitive diversity in team members has a wide range of benefits~\cite{vasilescu2015GenderTenureDiversity,pieterse2006SoftwareEngineeringTeam,capretz2010WhyWeNeed}. Furthermore, links between identity diversity and cognitive diversity~\cite{page2008difference,burnett2016GenderMagMethodEvaluating} further suggest that improving support for cognitive diversity may improve diversity in general.

However, LLMs do not fully support cognitive diversity, as indicated by the large experience and gender gaps in the adoption of LLM-based tools~\cite{draxler2023GenderAgeTechnology}. Similar indications have been found regarding gender, experience and problem-solving styles and LLMs in software engineering~\cite{russo2024NavigatingComplexityGenerative}. Prior work has identified several inequities in LLM-based tools for coding tasks. These include differences related to problem-solving styles in code understanding~\cite{nam2024,anderson2025LLMsAttemptsAdapt}, programming experience in code generation~\cite{nguyen2024,kazemitabaar2023StudyingEffectAI}, and gender in interactions with ChatGPT for coding~\cite{choudhuri2024HowFarAre}.

Despite many cases where LLM-based tools fail to support aspects of cognitive diversity in software engineering, addressing these inequities requires a better understanding of how cognitive diversity shapes developers’ needs during interaction. To this end, we conducted an exploratory study of developers’ interactions with GitHub Copilot chat to identify these needs and examine their relationship to cognitive diversity. We focus on experience and problem-solving styles, which prior work indicates are relevant to programming assistants.

We conceptualize interaction as distributions over observable \textit{interaction modes}, reflecting the possibility developers may switch between different patterns of interaction with GitHub Copilot chat over time, which is consistent with findings regarding interaction with GitHub Copilot inline suggestions~\cite{barke2023GroundedCopilotHow}. We take a broad interpretation of \textit{needs} to be any motivation, goal, desire, or preference, which allows us to inductively identify them by avoiding a rigid prior definition. Using this interpretation, we formulated the following research questions:

\vspace{3mm}
\boxRQ{I}{How do developers' exhibited interaction modes vary when using GitHub Copilot chat for code change tasks?}
\vspace{-1mm}
\boxRQ{II}{What needs drive developers' exhibited interaction modes?}
\vspace{-1mm}
\boxRQ{III}{How do developers' exhibited interaction modes relate to their experience profile and problem-solving styles, perceived ease of use and usefulness, and cognitive load?}

To answer these research questions, we conducted a think aloud study with $27$ participants completing code-change tasks using GitHub Copilot Chat (hereafter referred to as Copilot). Combining topic modeling of interaction data with qualitative analysis of think aloud data and post-study retrospective interview transcripts, we identify five interaction modes, ten underlying needs, and links to cognitive diversity. The resulting conceptual model supports reasoning about developer-programming assistant interactions and informs future work on improving inclusivity.
\section{Background \& Related Work}
\subsection{Cognitive Diversity}
Cognitive diversity is a broad concept, spanning trait-like aspects such as cognitive style, mental ability, and personality, as well as acquired aspects such as beliefs, educational background, and experience~\cite{mello2015CognitiveDiversityTeams}. Cognition affects a range of factors like decision making, teamwork, and collaboration~\cite{mello2015CognitiveDiversityTeams}, and is heavily involved in software engineering activities~\cite{fagerholm2022CognitionSoftwareEngineering}. 

The GenderMag method introduced an operationalization of cognitive diversity to the field of software engineering, by identifying five facets of cognitive diversity (self-efficacy, motivation, learning style, information processing style, and attitude towards risk) related to how people use software tools for problem-solving~\cite{burnett2016GenderMagMethodEvaluating}. The facets were selected to highlight gender differences in problem-solving and to help developers of problem-solving tools be more aware of how inclusive their tools are to gender.

\subsection{Cognitive Diversity in Developer-AI Interaction}
Several studies have studied LLM-based assistance in SE through the lens of cognitive diversity. Anderson et al.~\cite{anderson2025LLMsAttemptsAdapt} studied how LLM-generated code explanations were perceived by individuals with different problem-solving styles, finding inequities across all five.

Nam et al.~\cite{nam2024} designed an LLM-based information support assistant that allows interaction through both natural language and a set of buttons. They identified effects of the GenderMag information processing and learning style on developer interactions, and found novice developers were not as well supported as professionals.

Through a qualitative analysis of a survey among 100 software engineers, Russo~\cite{russo2024NavigatingComplexityGenerative} found indications that perceived ease of use may be influenced by Gendermag's learning styles and problem-solving approaches.

\subsection{Characterizations of Developer-AI Interaction}
There are many studies that aim to characterize developer-AI interactions and the drivers behind them. Treude and Gerosa~\cite{treude2025HowDevelopersInteract} propose a taxonomy of 11 overarching developer-AI interaction types, such as conversational assistance or auto-complete code suggestions. Barke et al.~\cite{barke2023GroundedCopilotHow} focuses on a single interaction type, by studying developers' interaction with the inline suggestion functionality of GitHub Copilot. They identified exploration and acceleration as two interaction modes that developers switch between, and characterized the conditions that trigger these modes and the strategies participants use while in a mode.

Recent work has increasingly examined how developers interact with conversational AI, and there are several characterizations of prompting patterns, triggering conditions, and usage patterns~\cite{nam2024,choudhuri2024HowFarAre,khojah2024,tie2024LLMsAreImperfect,shin2025PromptEngineeringFineTuning}. Some of these patterns are linked to developers' experience with conversational AI or with programming~\cite{nam2024,khojah2024,tie2024LLMsAreImperfect}, and there are hints that other individual factors play a role too~\cite{nam2024,tie2024LLMsAreImperfect,choudhuri2024HowFarAre}.

Although prior work addresses many of the concepts relevant to this study’s research questions, little is known about how they relate to one another, particularly the connections between patterns in interaction, underlying needs, and cognitive diversity.
\section{Methods}
\label{sec:methods}
\subsection{Participants}
We recruited professional software engineers (both employed and unemployed) and students enrolled in software engineering programs. All authors distributed the call for participation within their professional networks, while snowballing was employed by asking recruited participants to further redistribute the call. As identity diversity (e.g., age, gender, and nationality) has been linked to cognition~\cite{page2008difference,russo2022GenderDifferencesPersonality}, and cognitive diversity also encompasses factors such as educational and professional background, we aimed to maximize diversity across these aspects. To this end, our call for participation emphasized the relevance to inclusivity and encouraged anyone to join.

A total of 29 participants were recruited, of whom one was excluded due to a Copilot malfunction and another due to loss of interaction data, leaving 27 participants included. As demonstrated in Table \ref{tab:demographics}, we achieved reasonable diversity across most aspects, although gender diversity was limited with only 4 participants ($<15\%$) identifying as female.

\begin{table}[htbp]
    \centering
    \caption{Participant demographics}
    \vspace{-2mm}
    \label{tab:demographics}
    \begin{tabularx}{\linewidth}{p{3cm}X}
\toprule
\textbf{Age} & 18-24 (15), 25-34 (10), 35-44 (2) \\
\textbf{Gender} & Male (22), Female (4), Prefer not to say (1) \\
\textbf{Country} & Brazil (13), Netherlands (5), Romania (2), Russia (1), United Kingdom (1), Italy (1), India (1), Japan (1), China (1), Estonia (1) \\
\textbf{Professionals (18)} & Developer, full-stack (6), Academic researcher (3), Data scientist or machine learning specialist (2), Developer, back-end (2), Developer, front-end (2), Scientist (1), Developer, AI (1), Other, unrelated to software development (1) \\
\textbf{Students (9)} & Undergraduate, not first year (4), Graduate (4), Undergraduate, first year (1) \\
\bottomrule
\end{tabularx}
\end{table}
\subsection{Study Design}
\paragraph{Cognitive Profiles}
Cognitive profiles, composed of participants' experience profiles and problem-solving styles, were gathered through a pre-study survey. The experience profile included 5-point Likert scales for self-reported proficiency (No experience - Expert) in programming in general, JavaScript, React, web development, and the IDE Visual Studio Code, as well as frequency of use of AI programming assistants (Never - Almost always). We combined JavaScript, React, and web development proficiency using Principal Component Analysis (PCA), since these measures showed strong to very strong correlations. Because the first PCA component explained almost all (87.7\%) of the variance, it was selected as a single measure of domain experience.

To assess other aspects of cognition beyond experience, we employed the GenderMag questionnaire, which measures diversity in software-based problem-solving styles across five facets~\cite{hamid2024measure}. Although GenderMag was developed to highlight gender differences in cognition and does not encompass the full spectrum of cognitive diversity, the facets still reflect general variation in problem-solving styles. We decided to employ GenderMag since it is a theoretically informed and empirically validated framework widely adopted in software engineering research (e.g., \cite{anderson2025LLMsAttemptsAdapt,nam2024}), supporting construct validity and enabling comparison with prior research. As our analysis relies on correlations rather than group comparisons, we refrained from binning scores into binary variables and instead retained the original 9-point Likert composites to better reflect problem-solving styles as spectra. These spectra range from more ``Abi''-like problem-solving styles to more ``Tim''-like problem-solving styles, which represent facet values typically seen mostly in females or males, respectively~\cite{burnett2016GenderMagMethodEvaluating}.

\begin{figure}[htbp]
    \centering
    \includegraphics[width=\linewidth]{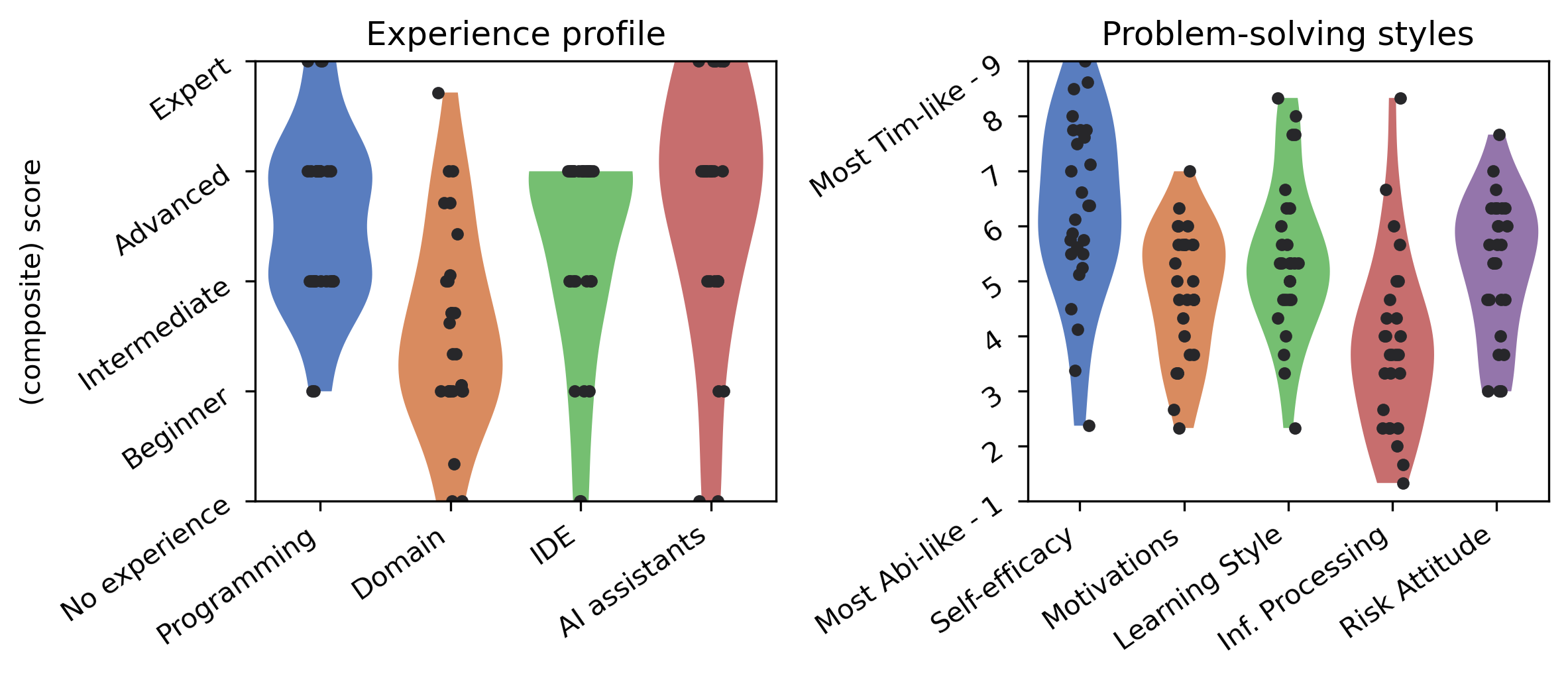}
    \vspace{-6mm}
    \caption{Distributions in participants' cognitive profiles, consisting of an experience profile and problem-solving styles.}
    \label{fig:cognitive_profiles}
\end{figure}

Figure \ref{fig:cognitive_profiles} shows broad variation in our participants' experience profiles and problem-solving styles, indicating substantial cognitive diversity in our sample.

\paragraph{Tasks}
To study how participants interact with Copilot, we constructed a set of code change tasks in a single codebase. We selected the React version of a popular demo project\footnote{\url{https://github.com/tastejs/todomvc}} to strike a balance between complexity: large enough to elicit realistic interactions while remaining manageable for participants of varying experience levels. 

Two of the authors, with respectively 13 and 2 years of professional software development experience, created 6 code change tasks of progressively increasing difficulty, asking participants to implement, modify, or debug a feature. Each task included a checklist of requirements and verification steps for participants to ensure their solutions were working and could be solved independently of one another. Participants had 40 minutes to complete tasks, and they were not required to finish all of them. The first three tasks were designed as practice to familiarize participants with the codebase and Copilot, while the final three were designed to be more realistic. Tasks and instructions were incrementally refined through a pilot study with one student and two professional developers to achieve a balanced level of difficulty. Participants rated the full tasks they attempted as highly to extremely realistic and progressively more difficult (5-point Likert scale). All participants attempted the three practice tasks, while 24, 15, and 9 participants attempted the first, second, and third full tasks, respectively.

This study design introduces order effects and means missing data is not random, as later tasks were only completed by participants who progressed through earlier ones. These limitations were deemed acceptable because our analysis focuses on participant-level interaction patterns rather than task-level differences. Our quantitative analysis further accounts for potential biases introduced by this setup.

\paragraph{GitHub Copilot Chat}
While completing the tasks, participants were allowed access to the chat functionality of GitHub Copilot. This is an in-IDE programming assistant, with access to the codebase. Participants used the most recent version of GitHub Copilot chat at the time of their session, which took place between March 27 and May 28, 2025. Although Copilot received minor updates during this period, we monitored sessions to ensure its behavior remained consistent. After observing a major update that introduced behavioral changes and a bug during a later session on June 17, we stopped data collection and excluded that participant.

In the versions used in this study, Copilot had access to the currently open file by default. Users could contextualize their questions to the entire codebase by including a contextualization command in their prompt. Generated code suggestions appeared only in the chat interface, and could either be copied manually or applied through a Git-style diff viewer accessed via an ``apply'' button. Participants were asked not to enable non-chat Copilot features.

\paragraph{Study Protocol}
Three of the authors conducted sessions, which were conducted remotely through a Microsoft Teams call between one researcher and the participant. Participants were given access to an instance of GitHub Codespaces\footnote{\url{https://github.com/features/codespaces}}, a browser-based version of Visual Studio Code. The environment was preconfigured by the authors to install required extensions and open directly into the codebase.

Sessions followed the think aloud method~~\cite{vansomeren1994ThinkAloudMethod}, with participants asked to narrate their thoughts as they worked. This method allows for insights into peoples cognitive processes as they problem-solve~\cite{vansomeren1994ThinkAloudMethod}. The researcher first explained this method to the participant and then introduced them to Copilot, demonstrating the contextualization functionality and command. Participants were then asked to attempt to run the application themselves, with permission to use Copilot. If unsuccessful after five minutes, guidance was provided.

Participants began with the three practice tasks to familiarize themselves with the codebase and Copilot before proceeding to the three full tasks. The combined practice and full task phase lasted up to 40 minutes. During the session, the researcher reminded participants to think aloud when necessary. Participants were instructed not to use external support tools other than Copilot. However, they were allowed to use a search engine if they became completely stuck and Copilot did not help. No participants used this option.

After the tasks, participants completed a post-study survey. This was followed by a ~25-minute semi-structured post-study interview exploring how Copilot influenced navigation through the project, understanding of the codebase, and problem solving, as well as aspects of the interaction participants liked or disliked. The interview also allowed the researcher to follow up on observations made during the task phase. By combining the think aloud method with retrospection, we aimed to gain richer insights and fill in gaps~\cite{vansomeren1994ThinkAloudMethod}.

\paragraph{Data Collection}
Participants' contact details, demographics, experience profile, and problem solving styles were gathered in a pre-study survey, which we distribution in the recruitment call. The post-study survey
gathered participants perceptions regarding the difficulty and realism of the tasks, as well as perceived usefulness and ease of use using the Technology Acceptance Model (TAM)~\cite{davis1989}, and perceived cognitive load using the Task Load Index (NASA-TLX) questionnaire~\cite{hartDevelopmentNASATLXTask1988}. 

During sessions, participants' screens and audio were recorded via Microsoft Teams, which also transcribed the audio. The codebase after the participants' modifications and Copilot chat log were exported after the session. Telemetry data, including file changes and open/close events, was collected via an existing extension\footnote{\url{https://github.com/educational-technology-collective/vscode-telemetry}} that we adapted to our needs. Further interactions with the Copilot interface, such as pressing the `apply' button, were annotated manually afterwards by watching the screen recordings. Audio transcripts and telemetry were manually corrected, after which raw audio and video data were deleted. Since we offered participants the option to conduct the sessions in Brazilian Portuguese, we automatically translated the audio transcriptions and chat logs to English using the GPT-4o LLM\footnote{\url{https://openai.com/index/hello-gpt-4o/}} via the OpenAI API. GPT-4o was prompted to translate word-for-word, leaving intact mistakes, sounds, and punctuation. All translated transcripts were manually validated, with all ambiguities being resolved by at least one native-speaking author. Audio transcripts, telemetry data, and chat logs were combined into single transcripts for further qualitative analysis.

The study protocol, task materials, adapted codebase and telemetry extension, anonymized transcripts, survey responses, and chat logs are available in the replication package~\cite{richards_2026_20734142}, in English and Brazilian Portuguese where appropriate. The study material was translated to Brazilian Portuguese using GPT-4o similar to transcript translation, with at least one native-speaking author manually adapting and improving the translations.

When preparing and executing this study, we followed the guidelines set by Radboud University's Research Ethics Committee and received its approval to conduct the study.
\subsection{Characterizing Interaction Modes}
To identify and characterize interaction modes (which we previously defined as observable patterns in developers' interactions over brief periods of activity) and to answer \refRQ{I}, we employed topic modeling, a set of unsupervised machine learning techniques from natural language processing (NLP). However, our aim was not to find patterns among words directly. Instead, we annotated each prompt on several axes using a hybrid of inductive and deductive open coding~\cite{corbin2015BasicsQualitativeResearch}: intent, additional information to support intent, contextualization method, whether the correct file for the query was open, implementation method, and short-term strategies, such as whether the participant rephrased or recontextualized a previous prompt. The coding scheme was iteratively refined through discussions between two authors. The resulting annotations were then used as tokens for topic modeling.

We first performed topic modeling on individual prompts to reduce dimensionality and prevent within-prompt patterns from dominating interaction modes. We employed Latent Dirichlet Allocation (LDA), with each prompt representing a document and annotations as tokens to obtain a distribution over 16 topics (referred to as prompt types) for each prompt. The prompt types were given semantic names based on their significant annotations to aid interpretation in the next steps.

Next, we performed topic modeling over sequences of prompts, using a sliding window to capture participants switching between interaction modes within a task. We obtained the documents for topic modeling by summing the distributions of prompt types within a sliding window over the prompts for each participant-task combination in which Copilot was used. A similar approach to ours has been used to identify temporal activity patterns using LDA~\cite{huynh2008DiscoveryActivityPatterns}. We used non-negative matrix factorization (NMF) instead, since the features now represent summed probability distributions rather than counts. Prompts related to running the application were grouped into a separate ``pseudo-task'' to avoid combining them with task-specific interactions.

We performed hyperparameter optimization for both LDA and NFM. For LDA, we optimized with respect to the number of significant annotations (weight over a threshold of 5\%) in at least one prompt type, reaching a local maximum at 16 prompt types. For NFM, we optimized by balancing low average window-mode entropy with high mode similarity, resulting in five interaction modes and a window size of three as a reasonable trade-off.

Prompt-mode distributions were obtained by averaging the mode distributions of all windows containing a given prompt. Mode distributions per participant-task observation were then obtained by averaging across all the prompts within a task. Finally, we sought to find the distribution over interaction modes for each participant. To reduce task-specific effects and account for order biases and non-random missing data, we adjusted the distributions of participant-task modes. While mixed-effects regression would be statistically preferable, the exploratory nature of the study and the limited sample size led us to use a simpler approach. For each task, we computed the average mode distribution, divided each participant's task distribution by the task average to remove task effects, and then averaged across tasks to obtain a participant-level mode distribution.

To facilitate interpretation of the interaction modes, we computed descriptive statistics at the prompt and task levels. For each mode, expected values were calculated as weighted averages of the relevant statistics, using mode probabilities as weights. First, we computed the expected number of characters per prompt, the expected number of characters when omitting copied code and task instructions, and the expected number of prompts per task. We also selected examples of prompts for the distinctive prompt types contributing to each mode.
\subsection{Identifying Developer Needs Driving Interaction Modes}
To identify the needs driving interaction modes and answer \refRQ{II}, we conducted a qualitative analysis of the transcripts, drawing on tools from grounded theory (open and axial coding)~\cite{corbin2015BasicsQualitativeResearch}. Although borrowing from grounded theory, we did not attempt to develop a formal grounded theory or employ theoretical sampling.

Starting with open coding, we inductively annotated fine-grained codes reflecting preferences, desires, behavioral patterns, and participants' articulated reasoning and motivations connecting these aspects. The first two transcripts were coded independently by two authors, after which codes were compared, discussed, and an initial coding scheme was agreed upon. The other transcripts were coded by one author, with codes being iteratively refined through constant comparison. In the following axial coding phase, we grouped these fine-grained codes into higher-level categories and relationships. As this qualitative analysis was meant to be explanatory regarding participants' interaction modes, we used deductive coding to guide the grouping of behavioral patterns into categories corresponding to the interaction modes. Other categories were discovered inductively, and reflected individual factors, contextual factors, underlying needs, and relationships between these categories and the interaction modes. During axial coding, two authors held repeated discussions to refine categories and resolve ambiguities. After participant 17 no new axial codes were identified, demonstrating saturation, with subsequent participants contributing only to refinement of existing categories rather than new ones.
\subsection{Identifying and Exploring Relationships Between Interaction Modes and Developer Cognitive Profiles and Perceptions}
To answer \refRQ{III}, we combined quantitative analysis with insights gained from our qualitative coding. First, we correlated participants' interaction mode distributions with their experience profiles, problem-solving styles, TAM scores, and TLX scores. Since the distributions are compositional data, they are subject to a constant-sum constraint and cannot be directly used for correlation~\cite{pawlowsky-glahnLectureNotesCompositional}. To remove this constraint, we applied a centered log-ratio (CLR) transformation before correlating. Because the CLR-transformed distributions and some profile and outcome variables were not normally distributed (as determined by Shapiro-Wilk tests), we employed Spearman's rank correlation. Next, we interpreted the identified correlations by relating them to the insights from our qualitative analysis.
\section{Results}
\subsection{Interaction Modes}
\label{sec:modes}

\begin{figure}[htbp]
    \centering
    \includegraphics[width=\linewidth]{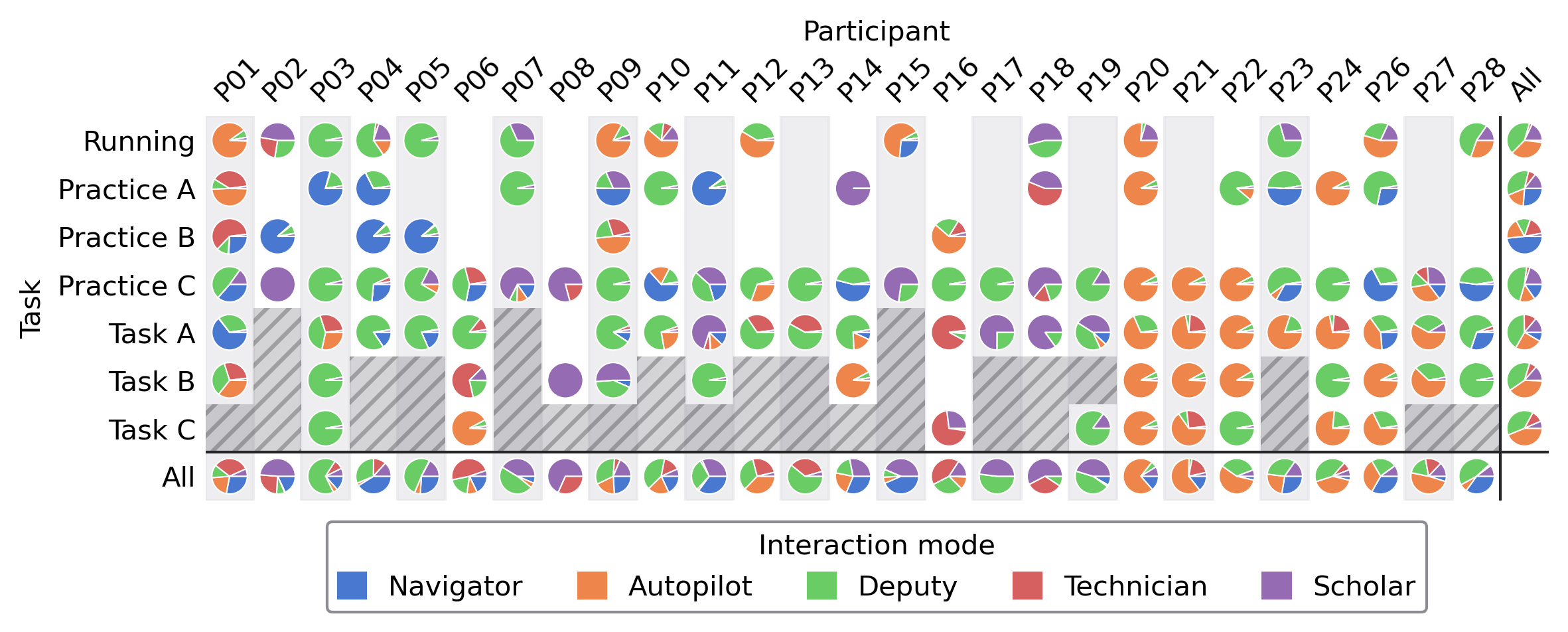}
    \vspace{-6mm}
    \caption{Interaction mode distributions per participant-task combination. Average task-mode distributions are shown in the right column, the bottom row shows task-adjusted participant-level mode distributions. Tasks not attempted by participants are hatched.}
    \label{fig:styles}
\end{figure}

Figure \ref{fig:styles} shows participants' interaction modes per task, and overall. Copilot was not used at all in only 2 of the 54 full task attempts. Task-adjusted participant-level mode distributions show clear variation in composition, reflecting high diversity in how participants interacted with Copilot.

\begin{figure}[htbp]
    \centering
    \includegraphics[width=\linewidth]{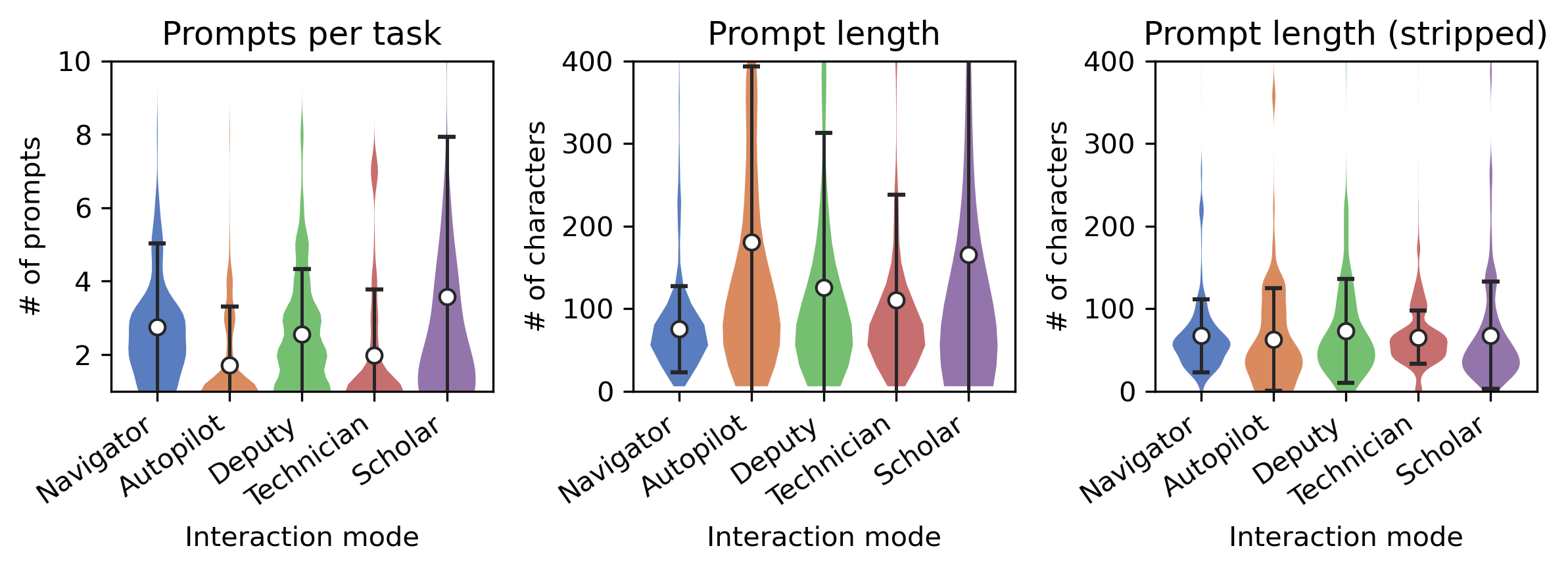}
    \vspace{-6mm}
    \caption{Expected distributions per mode of the number of prompts per task, prompt length, and prompt length stripped of copy/pasted code and task instructions.}
    \label{fig:mode_stats}
\end{figure}

Below, we describe the five interaction modes that we identified, and give examples of the prompt types characteristic of each.

\medskip
\paragraph{Navigator Mode}\mbox{}\vspace{1mm}
\boxExample{Locate}{@workspace Where is the red box defined when I dont have any todos yet}{P03}

When in Navigator mode, participants used Copilot to navigate the codebase. They would attempt to identify either high-level features (e.g., a button) or low-level code features (e.g., a function). Prompts would typically be sent while looking at a file not relevant to their query. Notably, participants would often omit the contextualization command and instead instruct Copilot to contextualize through natural language (e.g., ``find in my files [...]'', P04), which usually led to uncontextualized responses. Navigator mode was frequently used in tasks, either alone or together with Deputy mode, and was more often used in practice tasks (Figure \ref{fig:styles}). Being used in tandem with Deputy mode may explain the relatively high expected number of prompts per task (Figure \ref{fig:mode_stats}).

\medskip
\paragraph{Autopilot Mode}\mbox{}\vspace{1mm}
\boxExample{Solve, end-to-end}{\#codebase [copied task text] Fix or reimplement the navigation and implement the other requirements}{P20}
\vspace{-1mm}
\boxExample{Debug (end-to-end)}{@workspace It's still not working. I'm typing in the text input and clicking to reload the page, and when it reloads, no saved draft appears.}{P22}

In Autopilot mode, participants would attempt to use Copilot to solve their problem completely automatically. The related prompt types (Solve/Debug end-to-end) are characterized by prompting with copied task instructions or high-level goal descriptions, often using the contextualization command outside the relevant file, and implementing using the `apply' button. The low expected number of prompts per task (Figure \ref{fig:mode_stats}) and Autopilot's tendency to dominate the task-mode distribution when employed (Figure \ref{fig:styles}) reinforce the interpretation of this mode as a ``one and done'' strategy. Autopilot mode was used more often in the full tasks than in the practice tasks.

\medskip
\paragraph{Deputy Mode}\mbox{}\vspace{1mm}
\boxExample{Solve, incremental}{Change the code to store the input field text every 2 seconds to local state, and restore it when I initialize if any text was entered}{P03}
\vspace{-1mm}
\boxExample{Debug, targeted}{Is the todoReducer function complete? Specifically the LOAD\_STATE case?}{P09}
\vspace{-1mm}
\boxExample{Explain response}{Can you just explain to me what the error was and what your changes were in brief. I am not an expert in Javascript nor have I written this code myself and just looking to fix this issue, but also want to understand a bit}{P14}

While in Deputy mode, participants used Copilot to incrementally and collaboratively work towards solutions. This mode combines the largest number of prompt types, of which three distinct examples are shown above. Participants would often use this mode while already in (one of) the files relevant to their problem, and frequently refer back to previous interactions in the chat. Generated code was usually implemented by copying/pasting (parts of) snippets. The incremental nature of this mode is supported by the relatively high number of prompts per task. Notably, participants often would not contextualize their prompts when needed. However, users were usually able to detect this and re-contextualize their previous prompt accordingly. Participants asking Copilot to explain its responses reinforces that this mode collaborative. Figure \ref{fig:styles} shows Deputy is the dominating interaction mode, and is used throughout all tasks.

\medskip
\paragraph{Technician Mode}\mbox{}\vspace{1mm}
\boxExample{Implement, low-level}{use debounce to handle the input change}{P06}
\vspace{-1mm}
\boxExample{Implement, mid-level}{I need the "todos" localStorage to be loaded and feed the variables.}{P16}

Technician mode is characterized by participants using Copilot to generate code for them for granular changes, through either low- or mid-level instructions closely related to the code rather than full descriptions of their goals. Generated code is typically manually implemented (by looking at it and typing it in the codebase themselves), or copy/pasted in small snippets. The low expected length of prompts (Figure \ref{fig:mode_stats}) supports the interpretation of using Technician mode for granular and concrete changes. Technician mode seems to often be used in small amounts in participants' tasks, in combination with other modes (Figure \ref{fig:styles}).

\medskip
\paragraph{Scholar Mode}\mbox{}\vspace{1mm}
\boxExample{Explain code}{What is this React code doing? "[copied code]"}{P18}
\vspace{-1mm}
\boxExample{Information Support}{Javascript lexographical sort of strings}{P07}

Participants in the Scholar mode used Copilot to explain code or provide information support. They would typically phrase their prompts using low-level terms close to the code. Usually, this mode was exhibited while already looking at the file relevant to their prompt, or when their request for information was so general that no file was needed for contextualization. Still, participants would often redundantly copy/paste code into Copilot, as reflected in the large difference between full and stripped prompt length (Figure \ref{fig:mode_stats}). Participants tended to send many prompts to Copilot in this mode. When used, Scholar mode seems to either dominate the strategy for a task or co-occur with the Deputy and Technician modes (Figure \ref{fig:styles}).

\vspace{1mm}
\boxSummary{
We identified and characterized 5 interaction modes exhibited by developers through topic modeling of their prompts, answering \refRQ{I}.
}
\subsection{Needs Driving Interaction Modes}
\label{sec:needs}
\begin{figure}[hptb]
    \centering
    \subfloat[Needs and relationships with individual and contextual factors.]{%
        \includegraphics[width=\linewidth]{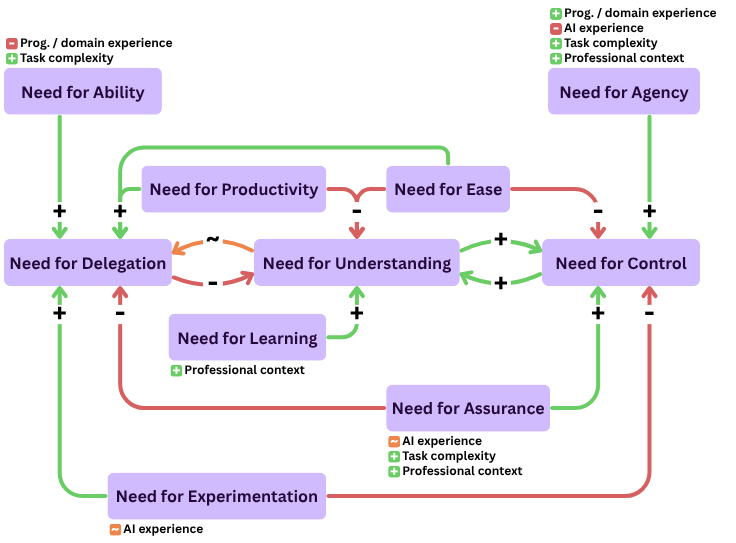}
        \label{fig:needs}%
    }
    
    \subfloat[Interaction modes and relationships with needs and individual and contextual factors. Only needs directly linked to interaction modes are shown.]{%
        \includegraphics[width=\linewidth]{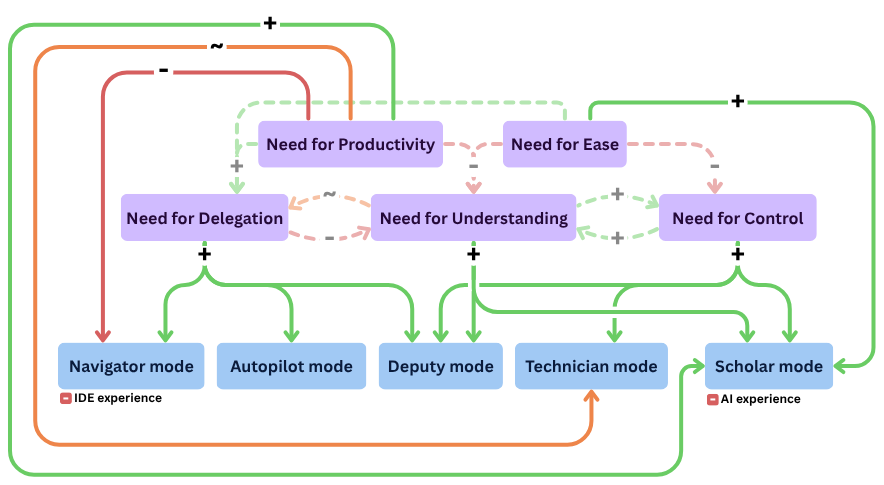}
        \label{fig:modes_links}%
    }
    
    \caption{Developer needs (purple) and interaction modes (blue) when interacting with GitHub Copilot chat, and relationships between these needs, interaction modes, and individual and contextual factors. Relationships may be positive ($+$), negative ($-$), or mixed ($\sim$).}
    \label{fig:needs_modes_links}
\end{figure}

In our open and axial coding of think aloud and interview transcriptions, we identified 10 core needs relevant to interactions with Copilot and their relationships (Figure \ref{fig:needs}). Relationships between needs can be interpreted in either of two ways, e.g.: \begin{enumerate*}[label=\roman*)]
    \item a need for ability induces or increases a need for delegation, or
    \item delegation supports ability
\end{enumerate*}. We also identified relationships between these needs and individual and contextual factors (Figure \ref{fig:needs}), as well as earlier characterized interaction modes (Figure \ref{fig:modes_links}). Note that these links reflect participants' perceptions. Additional links, such as a bidirectional negative relationship between the need for delegation and control, are intuitively plausible but are not directly grounded in our thinking-aloud and interview data.

Together, the relationships between needs, modes, and individual and contextual factors form a conceptual model of how developer's interactions with Copilot are shaped (Figure \ref{fig:needs_modes_links}). This model shows the complex links between needs and interaction. For example, the need for ease may induce the need for delegation and therefore the usage of deputy mode, but may also deter the usage of deputy mode by reducing the need for control.

\paragraph{Need for Ability} The need for ability arises from the gap between the complexity of a task and developers' capacity to solve it themselves. When in need of ability, developers increasingly delegate to Copilot.

\paragraph{Need for Agency} Agency covers a range of needs related to wanting to be self-reliant. Professional contexts heighten the need to demonstrate competency through agency, and unfamiliarity with AI assistants can also increase the need for self-reliance. The need for agency increases with task complexity, as developers like to self-challenge and enjoy problem-solving. Developers also feel they should be self-reliant once they have experience in programming or the relevant domain. Pride, self-challenging, self-reliance, and a desire for perceived authorship decrease delegation and increase control over solutions.

\paragraph{Need for Experimentation} Some developers desire or are open to experimenting with Copilot. This can relate to the degree to which they delegate or the release of control over Copilot suggestions. Having preconceptions about Copilot's capabilities from (lack of) prior experience with AI assistants can limit this need.

\paragraph{Need for Assurance} Assurance covers the gap between developers' need to have trust in the adequacy of solutions to their problems, and the amount of trust they have in Copilot's support. Professional contexts require greater trust, and expectations of Copilot's performance decrease for more complex tasks and when developers have preconceptions stemming from (lack of) prior experience with AI assistants. When assurance is needed, developers delegate less to Copilot and take more control of solutions.

\paragraph{Need for Productivity} Productivity encapsulates the desire to save time or get more done in the same amount of time. It reduces the desire to fully understand Copilot suggestions, as this takes time and is supported by delegation, speeding up tasks. Productivity may be hindered by the time required to write prompts and contextualize responses, when developers can use file search instead of Copilot to locate (Navigator) or write code manually instead of using Copilot (Technician). Still,  using Copilot to write code (Technician) or for information support (Scholar) can increase productivity.

\paragraph{Need for Ease} Closely related to the need for productivity, ease covers wanting to save physical and mental effort, and avoid difficulties. This reduces the desire for control and understanding, as thinking about implementation details, manually implementing Copilot suggestions, and understanding the codebase and solutions takes effort. Instead, developers delegate more to Copilot to increase ease of use. Specifically, Copilot saves effort over using a web browser when used for information support (Scholar) by being in-IDE and automatically finding relevant knowledge.

\paragraph{Need for Learning} In some cases, developers may desire or need to learn about the codebase or develop their skills. This need arises in a professional context and requires an understanding of the codebase and problem-solving skills.

\paragraph{Need for Understanding} A need for understanding the codebase, Copilot suggestions, and solutions to problems is induced by needing to learn or to have more control. Vice versa, control also improves understanding when developers try to create their own solutions. Understanding is not needed when the goal is just to be productive or save effort, and when delegating to Copilot. Conversely, delegating may decrease understanding when relied on too much, but also increase understanding by demonstrating strategies for solutions. Asking Copilot to explain code (Scholar) or its own responses (Deputy) supports understanding.

\paragraph{Need for Delegation} Delegation involves using Copilot more, and the degree to which problem-solving is outsourced. It is increased by the need for ability, productivity, and ease, and impacted by the need for understanding. Developers may delegate more when experimenting, and less when they prefer to self-rely or need more assurance on Copilot's performance. Using Copilot to locate where to work on a problem (Navigator), solve entire tasks (Autopilot), and work collaboratively by asking for advice and incrementally solving tasks (Deputy) supports delegation. Although Technician mode involves consulting Copilot as well, the high level of control it entails makes it unsuitable for delegation.

\paragraph{Need for Control} refers to the developer being able to decide how to solve and implement a problem. This refers to solving problems manually and using Copilot less, or using Copilot in targeted ways. The need for control arises from a need for assurance or understanding of the solution, and decreases when developers are experimenting with Copilot or want to save effort. Control is supported by manually setting up solutions and using Copilot to fill in the gaps (Technician) or provide information support (Scholar), manually setting prompt context to improve responses (Technician, Scholar), providing low-level instructions and manually implementing suggestions (Technician) or decomposing problems into steps for Copilot, verifying suggestions, and incrementally building towards a solution (Deputy).

\vspace{1mm}
\boxSummary{
We identified 10 core needs driving developers' interactions along with links to our interaction modes, answering \refRQ{II}. Identifying possible links between these needs and individual and contextual factors builds towards \refRQ{III}.
}
\subsection{Relationships Between Interaction Modes and Developer Cognitive Profiles and Perceptions}
\label{sec:relationships}

\begin{figure}[htbp]
    \centering
    \includegraphics[width=\linewidth]{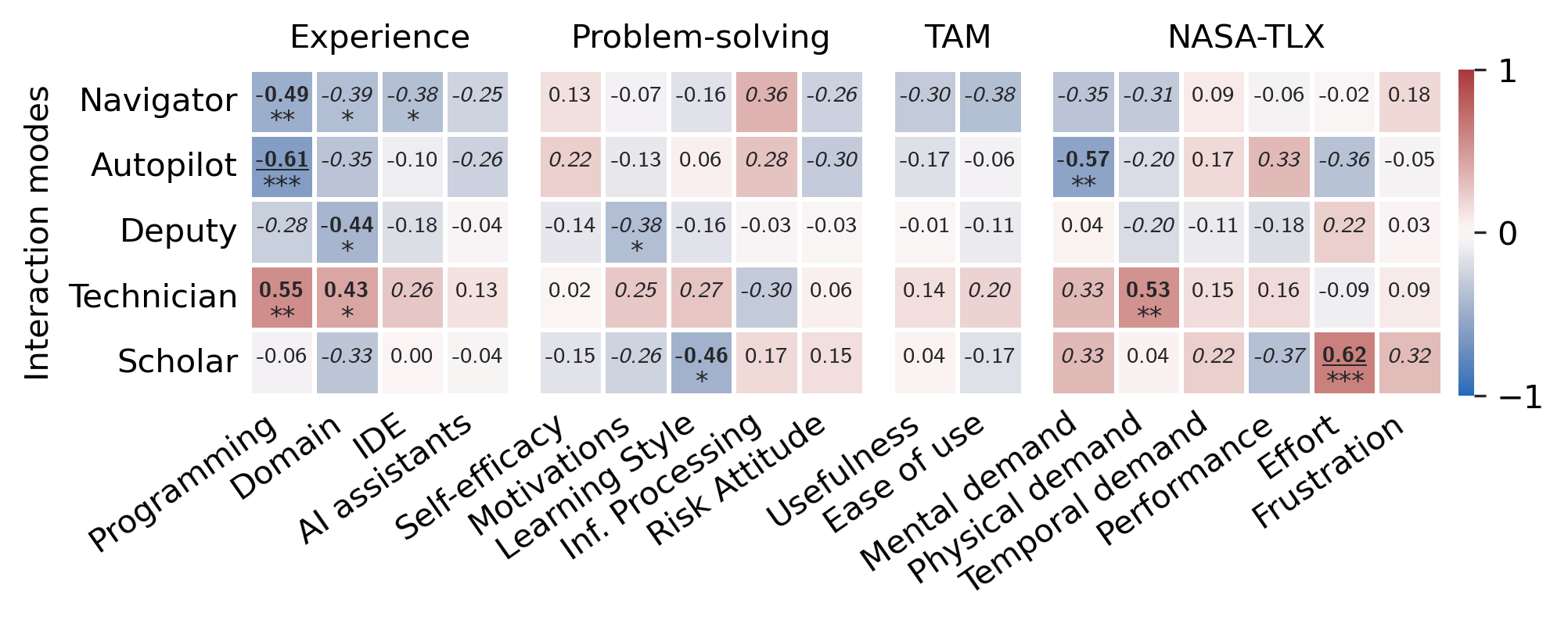}

    \vspace{-3mm}
    \caption{Spearman rank correlations between interaction modes and experience profile, problem-solving styles, perceived usefulness and ease of use, and cognitive load. Strong correlations ($|r| \geq 0.6$) are shown in bold and underlined, medium ($|r| \geq 0.4$) in bold, and weak ($|r| \geq 0.2$) in italics. Asterisks indicate significance: * $p < 0.05$, ** $p < 0.01$, *** $p < 0.01$. }
    \label{fig:modes_profiles_outcomes}
\end{figure}

Figure \ref{fig:modes_profiles_outcomes} shows several significant weak to strong correlations between interaction modes and participants' profiles and perceptions. Given the exploratory nature of this study and the limited sample size, p-values were not corrected for multiple comparisons. Below, we interpret these correlations in the context of the conceptual model we developed (Section \ref{sec:needs}). However, it should be noted that causal links are not claimed, and explanations are open to interpretation.

\paragraph{Experience Profile}
Programming and domain experience correlate with interaction modes: higher experience is associated with more Technician use and less Navigator, Autopilot, and Deputy. This aligns with Figure \ref{fig:needs_modes_links}, where experience reduces the need for ability and increases agency, shifting from delegation toward control. IDE experience is negatively correlated with Navigator mode. Some participants noted that file search is a good alternative that less experienced IDE users may not know about, as they were occasionally observed using Copilot to locate exact strings in the codebase.

\paragraph{Problem-solving styles}
The weakly positive though insignificant correlation between self-efficacy and Autopilot suggests that higher self-efficacy reduces need for assurance while supporting experimentation with Copilot, as developers trust their ability to address problems arising from delegation.

We encountered several weak correlations between problem-solving styles and interaction modes, although most are not statistically significant. Intuitively, one might expect high motivation scores (using technology for its own sake) to result in greater use of Autopilot mode, driven by participants' need to experiment and ``push Copilot to its limit'' to see what it can accomplish. Instead, motivation was positively correlated with Technician mode and negatively with Scholar and Deputy modes, although only the latter is statistically significant. This suggests motivation is only relevant to controlled interaction modes (Figure \ref{fig:needs_modes_links}). While Deputy and Scholar modes support participants’ needs for understanding, and delegation in the case of Deputy, Technician mode more directly reflects using the technology for its own sake.

The lack of positive correlations between a `tinkering' learning style (higher score) and Navigator and Autopilot mode, suggests that tinkering does not correspond to our need for experimentation (with Copilot) as one might expect. Rather, the medium negative correlation with Scholar mode and a weak positive (though statistically insignificant) correlation with Technician mode suggests it relates to using Copilot to tinker with participants' own solutions, rather than to using Copilot for structured guidance.

A selective information processing style (higher score) correlates positively with Navigator and Autopilot modes and negatively with Technician mode. Although care should be taken in interpreting this insignificant correlations, this may reflect selective processors’ tendency to delegate to Copilot to obtain an acceptable solution, whereas comprehensive processors may seek to understand possible solutions thoroughly and therefore rely more on controlled use through Technician mode. However, through the same reasoning it is not clear why no negative correlation with Scholar mode is found.

Intuitively, a more risk-tolerant attitude (a higher score) implies a lower need for assurance and thus greater delegation, leading to increased use of Autopilot and Navigator modes. However, Figure \ref{fig:modes_profiles_outcomes} shows the opposite, although not significantly. One explanation may be that risk tolerance in this case does not relate to using Copilot and failing, but instead to failing to create one's own solution to tasks. This suggests a positive relationship between the need for assurance and delegation, which is plausible as delegating to Copilot seemed to be a reliable strategy for participants in our study.

\paragraph{Perceived Usefulness and Ease of Use}
There were no significant correlations between perceived usefulness and ease of use and interaction modes, although Navigator mode was weakly negatively correlated with both measures. While this is not explained by Figure \ref{fig:modes_links}, it may relate to our earlier observation that participants expected Copilot to automatically contextualize responses in Navigator mode when asked to locate something, which it usually did not. Ease of use weakly positively correlates with Technician mode, suggesting that participants found this mode of interaction less effortful. However, our conceptual model shows a negative link between the need for ease and this mode (Figure \ref{fig:modes_links}). This supports that our need for ease relates to the task at hand, while TAM's ease of use relates to the effort in using Copilot.

\paragraph{Perceived Cognitive Load}
Autopilot mode positively correlated with perceived performance and negatively with effort and mental demand, although only the latter is significant. This is consistent with our finding that ease and productivity are supported by delegation.

Technician mode is associated with significant increased physical demands. This may be explained by participants implementing suggested changes either manually or by copying/pasting small snippets one at a time.

Although Scholar mode supported productivity and ease compared to traditional information support (Figure \ref{fig:modes_links}), its significant strong correlation with higher perceived effort suggest that this gain in ease does not compare to that of other modes. Instead, in this mode, the need for control or understanding may dominate the need for ease. The insignificant correlations with mental demand, temporal demand, and frustration, and the negative correlation with productivity support this notion.

\vspace{1mm}
\boxSummary{
We used our conceptual model to explore correlations between participants' interaction modes and their cognitive profiles and perceptions, answering \refRQ{III}. 
}
\section{Threats to Validity}
\textbf{Internal validity} is affected by our methodological and interpretive decisions. First, topic modeling requires choices such as hyperparameter settings and interpretation of identified topics. We mitigated this by using default values for established parameters, systematically optimizing others, and interpreting topics based on descriptive statistics. Inductive coding in our qualitative analysis is also subject to researcher interpretation, which we mitigated by involving multiple authors throughout the coding process, including independent coding, iterative discussions of the coding scheme, and the resolution of ambiguities.

\textbf{External validity} in our study is affected by the risk of overgeneralization. We attempted to mitigate this risk by recruiting a diverse participant sample, although we did not achieve substantial gender diversity. Additionally, the cognitive factors we examined may interact in complex ways beyond the scope of our exploratory study design. As a result, our findings should not be interpreted as universal patterns across developers. To support careful interpretation, we transparently report participant characteristics and avoid broad claims about individual differences, emphasizing instead that developers may exhibit diverse and unique needs. Additionally, it should be noted that this research is exploratory, and that the conceptual model has not been externally validated. Because harder tasks were attempted by less participants, the findings for these tasks were supported by less observations, raising the need for additional validation in real-life development scenarios.

\textbf{Construct validity} was addressed by clearly defining interaction modes (in line with prior research~\cite{barke2023GroundedCopilotHow}) and needs. The problem-solving styles we assessed were based on the empirically validated GenderMag facets~\cite{burnett2016GenderMagMethodEvaluating}. We further examined the consistency of our experience measures by confirming that self-reported proficiency correlated moderately to strongly with years of experience. Variance inflation factors (VIFs) for experience measures (1.00 to 1.65) and problem-solving styles (1.22 to 1.86) indicated negligible multicollinearity, suggesting that the chosen metrics capture distinct constructs.
\section{Discussion}
Our findings include a conceptual model of needs underlying developers' interactions with GitHub Copilot chat (Section \ref{sec:needs}), showing how needs can reinforce or compete with each other in shaping developers' behavior. Through this model, we explored how diversity in experience and problem-solving styles relates to developers' interactions (Section \ref{sec:relationships}).

Due to the exploratory nature of this study, we do not claim definitive interpretations of these relationships. For instance, a positive correlation between a mode and effort could indicate that using the mode increases effort, or that participants select the mode because conserving effort is not a priority. Determining causal relationships through controlled experiments is left to future work. Nevertheless, our conceptual model provides a basis for reasoning about developers' exhibited interaction modes and the underlying needs they entail, in relation to contextual and individual factors. It may guide researchers and designers of conversational programming assistants in explaining observed behavior and generating hypotheses about factors shaping interaction.

Treude and Gerosa~\cite{treude2025HowDevelopersInteract}
suggest researchers to find interaction types to best strike a balance between trust, automation, control, productivity, and cognitive burden. Our model extends this perspective by showing that these needs are not fixed. Rather, they vary with cognitive and contextual factors, leading developers to adopt different interaction behaviors in different situations. Therefore, future research should consider that improving interaction may require adapting to individual developers and contexts, rather than relying on one-size-fits-all solutions when designing and evaluating LLM-based support.

This work provides useful insights for all three human-AI experience research dimensions as identified by Sergeyuk et al.~\cite{sergeyuk2026HumanAIExperienceIntegrated}. Research into the \textit{impact} of LLM-based developer assistance should take into account the many competing needs driving developer behavior, possibly obscuring the effects of interest. When studying the \textit{design} of tools, researchers should take into account individual and contextual factors as described earlier. As noted by Sergeyuk et al.~\cite{sergeyuk2026HumanAIExperienceIntegrated} and supported by our findings, the ways developers prompt, verify, and debug when using LLM-based assistance are affected by human and contextual factors, showing their importance to research into \textit{quality} of these tools too. The conceptual model we present may be used by researchers to reason about these competing needs, differences in interaction, and underlying factors.

Our findings align with prior work, supporting the validity of our model. Nam et al.~\cite{nam2024} found that process-oriented learners interacted more with an information-support assistant, while tinkerers modified the code directly with less information support. This reinforces our interpretation that learning style affects the Scholar and Technician modes through the task itself, not Copilot usage.

Relating our modes to those identified by Barke et al.~\cite{barke2023GroundedCopilotHow}, Technician corresponds to acceleration, where developers retain control by decomposing tasks and focusing on granular code features. Our other modes align more with exploration: developers delegate problem-solving when they trust the tool or face unfamiliar tasks, or seek information when next steps are unclear. The imbalance between these characterizations (with Technician representing only 7\% of average task-mode distributions) likely stems from the different assistance studied: Copilot inline suggestions versus chat. Conversational interaction may not be most suitable for acceleration due to the effort in prompting and subsequent manual implementation of suggested changes from the chat, supported by the significant correlation between Technician mode and perceived physical effort. The low usage of Technician mode then suggests that developers adapt the support they seek to the perceived capabilities of the available tool, as reflected in participants' remarks.

\vspace{1mm}
\boxQuote{So, I would probably be asking him [...] to do it if he were in agent mode, but in text mode I only ask when there's really something I can't do without it, because, apparently the time to do and the time to ask is the same.}{P16, referring to using Copilot for implementing code}

The association between Technician mode and higher programming and domain experience, but also increased physical effort, supports Treude and Gerosa's notion that different interaction types may be more suited to particular user profiles~\cite{treude2025HowDevelopersInteract}. This raises the need for further research into cognitive diversity and LLM-based support beyond conversational interaction.

Other participants' comments similarly reflected how their use of Copilot was influenced by their perceptions of its capabilities. Overestimating Copilot can lead to disappointment or frustration, while not being aware of its capabilities can lead to underutilization. This highlights the importance of programming assistants clearly demonstrating what they can or cannot be used for, relating to the concept of ``affordances'' in HCI research~\cite{kaptelinin2012AffordancesHCIMediated} and basic guidelines of human-AI interaction~\cite{amershi2019}.

\vspace{1mm}
\boxQuote{Not necessarily what I was expecting.}{P07, after verbally instructing Copilot to search all files, which did not result in contextualization}
\vspace{-1mm}
\boxQuote{Oh, it can actually look at what the code base does, cool.}{P07, after asking a search-engine-like question}

We found several indications that participants ran into varying challenges and had different preferences based on their needs and exhibited modes. Some participants appreciated Copilot explaining its code suggestions, while others found them to hinder when Copilot was used to support productivity rather than understanding.

\vspace{1mm}
\boxQuote{And also I liked that he, when he gave the answer for each line, he gave a slight indication of why that line needed to be changed.}{P12}
\vspace{-1mm}
\boxQuote{I just needed a quick answer and then it sort of hindered because it just cluttered the rest of the information.}{P08}

Additionally, Copilot would often go beyond participants' explicitly stated intent, which participants appreciated or even expected when trying to be more productive or to delegate. Other participants even mentioned Copilot should have challenged them or asked for clarification when it thought their prompts did not align with their goals. On the other hand, when preferring control, participants opposed Copilot's reasoning about or extrapolating their intent.

\vspace{1mm}
\boxQuote{Sometimes it ends up doing more than it should, but. If the intention is to be productive. I'm using it a lot, and it's helping me a lot. I'm taking advantage.}{P16}
\vspace{-1mm}
\boxQuote{It didn't ask me like when I didn't say the workspace, it didn't bother asking.
Maybe your changes are somewhere else.}{P14}
\vspace{-1mm}
\boxQuote{[...] this is a bit of a step further that I don't even was. Not even plan on my list. [...] it's a bit too much.}{P05}

As we found that professional context and task complexity directly impact the need for ability, agency, assurance, and learning (Section \ref{sec:needs}), future work should more closely examine the effects of contextual factors on developers' needs in real-world settings. Additionally, the modes we identified reflect how developers used Copilot in the tasks; future research could study preferred forms of support instead. Future research should also investigate interaction modes beyond conversation, in addition to extending to other parts of the software development lifecycle such as planning and design. Finally, examining temporal dynamics between interaction modes may be insightful.

Based on our insights, we recommend \textit{organizations} to be aware of individual differences in needs of employees regarding programming assistants, providing support for varying workflows to foster a productive workplace. Additionally, collaborative workshops involving developers can facilitate the sharing of knowledge about the strengths and weaknesses of different forms of interaction with generative AI. We recommend \textit{tool builders} to provide granular support for different interaction modes, as we show these may affect preferences and encountered challenges.
\section{Conclusion}
In this study, we explored how cognitive diversity shapes developers’ interactions with GitHub Copilot Chat. Through a mixed-methods think aloud user study with 27 professional developers and students completing code change tasks, we identified five interaction modes exhibited by developers and ten underlying needs driving their interactions with Copilot, together forming a conceptual model of developer-Copilot interactions. Our analysis further links developers’ interaction mode distributions to differences in experience, problem-solving styles, perceived ease of use, perceived usefulness, and perceived cognitive load, highlighting how cognitive diversity may shape interactions with programming assistants.

Our contributions are a conceptual model that supports reasoning and hypothesis generation about how cognitive diversity shapes developer interactions, and a set of insights and recommendations for organizations, and researchers and designers of LLM-based programming assistants.

\textbf{Acknowledgments.} We thank the developers and students who participated for their contribution. Igor Wiese thanks CNPq (409359/2024-6, 444802/2024-0) and Fundação Araucária/Governo do Paraná (PRD2023361000043)

\textbf{AI Disclosure.}
AI was used for minor editing and grammar improvements, as well as automatic transcription and translation as outlined in Section \ref{sec:methods}.

\balance
\bibliographystyle{IEEEtran}
\bibliography{references}

@dataset{richards_2026_20734142,
  author       = {Richards, Jonan and
                  Alves de Oliveira, Bruno and
                  Oliveira, Iury and
                  Wiese, Igor and
                  Wessel, Mairieli},
  title        = {Replication package for "{{No Two Developers Think
                   Alike: How Problem-Solving Styles and Experience
                   Shape Needs in Conversational Interaction with
                   Copilot}}"
                  },
  month        = jun,
  year         = 2026,
  publisher    = {Zenodo},
  doi          = {10.5281/zenodo.20734142},
  url          = {https://doi.org/10.5281/zenodo.20734142},
}

@article{sergeyuk2026HumanAIExperienceIntegrated,
  title = {Human-{{AI}} Experience in Integrated Development Environments: A Systematic Literature Review},
  shorttitle = {Human-{{AI}} Experience in Integrated Development Environments},
  author = {Sergeyuk, Agnia and Zakharov, Ilya and Koshchenko, Ekaterina and Izadi, Maliheh},
  year = 2026,
  month = may,
  journal = {Empirical Software Engineering},
  volume = {31},
  number = {3},
  pages = {55},
  issn = {1382-3256, 1573-7616},
  doi = {10.1007/s10664-025-10793-0},
  urldate = {2026-06-15},
  langid = {english}
}

@inproceedings{treude2025HowDevelopersInteract,
  title = {How {{Developers Interact}} with {{AI}}: {{A Taxonomy}} of {{Human-AI Collaboration}} in {{Software Engineering}}},
  shorttitle = {How {{Developers Interact}} with {{AI}}},
  booktitle = {2025 {{IEEE}}/{{ACM Second International Conference}} on {{AI Foundation Models}} and {{Software Engineering}} ({{Forge}})},
  author = {Treude, Christoph and Gerosa, Marco A.},
  year = 2025,
  month = apr,
  pages = {236--240},
  doi = {10.1109/Forge66646.2025.00033},
  urldate = {2026-06-15}
}

@misc{anderson2025LLMsAttemptsAdapt,
    title = {An {LLM}'s {Attempts} to {Adapt} to {Diverse} {Software} {Engineers}' {Problem}-{Solving} {Styles}: {More} {Inclusive} \& {Equitable}?},
    shorttitle = {An {LLM}'s {Attempts} to {Adapt} to {Diverse} {Software} {Engineers}' {Problem}-{Solving} {Styles}},
    url = {http://arxiv.org/abs/2503.11018},
    doi = {10.48550/arXiv.2503.11018},
    urldate = {2025-11-04},
    publisher = {arXiv},
    author = {Anderson, Andrew and Piorkowski, David and Burnett, Margaret and Weisz, Justin},
    month = mar,
    year = {2025},
    note = {arXiv:2503.11018 [cs]},
}

@inproceedings{nam2024,
    address = {Lisbon Portugal},
    title = {Using an {LLM} to {Help} {With} {Code} {Understanding}},
    isbn = {979-8-4007-0217-4},
    url = {https://dl.acm.org/doi/10.1145/3597503.3639187},
    doi = {10.1145/3597503.3639187},
    language = {en},
    urldate = {2024-04-23},
    booktitle = {Proceedings of the {IEEE}/{ACM} 46th {International} {Conference} on {Software} {Engineering}},
    publisher = {ACM},
    author = {Nam, Daye and Macvean, Andrew and Hellendoorn, Vincent and Vasilescu, Bogdan and Myers, Brad},
    month = apr,
    year = {2024},
    pages = {1--13},
}

@article{barke2023GroundedCopilotHow,
  title = {Grounded {{Copilot}}: {{How Programmers Interact}} with {{Code-Generating Models}}},
  shorttitle = {Grounded {{Copilot}}},
  author = {Barke, Shraddha and James, Michael B. and Polikarpova, Nadia},
  year = 2023,
  month = apr,
  journal = {Proceedings of the ACM on Programming Languages},
  volume = {7},
  number = {OOPSLA1},
  pages = {78:85--78:111},
  issn = {2475-1421},
  doi = {10.1145/3586030},
  urldate = {2024-10-18}
}

@article{mello2015CognitiveDiversityTeams,
    title = {Cognitive {Diversity} in {Teams}: {A} {Multidisciplinary} {Review}},
    volume = {46},
    issn = {1046-4964, 1552-8278},
    shorttitle = {Cognitive {Diversity} in {Teams}},
    url = {https://journals.sagepub.com/doi/10.1177/1046496415602558},
    doi = {10.1177/1046496415602558},
    language = {en},
    number = {6},
    urldate = {2026-03-06},
    journal = {Small Group Research},
    author = {Mello, Abby L. and Rentsch, Joan R.},
    month = dec,
    year = {2015},
    pages = {623--658},
}

@article{burnett2016GenderMagMethodEvaluating,
    title = {{GenderMag}: {A} {Method} for {Evaluating} {Software}'s {Gender} {Inclusiveness}},
    volume = {28},
    issn = {0953-5438},
    shorttitle = {{GenderMag}},
    url = {https://doi.org/10.1093/iwc/iwv046},
    doi = {10.1093/iwc/iwv046},
    number = {6},
    urldate = {2024-06-27},
    journal = {Interacting with Computers},
    author = {Burnett, Margaret and Stumpf, Simone and Macbeth, Jamie and Makri, Stephann and Beckwith, Laura and Kwan, Irwin and Peters, Anicia and Jernigan, William},
    month = nov,
    year = {2016},
    pages = {760--787},
}

@book{page2008difference,
    title = {The difference: {How} the power of diversity creates better groups, firms, schools, and societies-new edition},
    publisher = {Princeton University Press},
    author = {Page, Scott},
    year = {2008},
}

@article{fagerholm2022CognitionSoftwareEngineering,
    title = {Cognition in {Software} {Engineering}: {A} {Taxonomy} and {Survey} of a {Half}-{Century} of {Research}},
    volume = {54},
    issn = {0360-0300, 1557-7341},
    shorttitle = {Cognition in {Software} {Engineering}},
    url = {https://dl.acm.org/doi/10.1145/3508359},
    doi = {10.1145/3508359},
    language = {en},
    number = {11s},
    urldate = {2024-09-06},
    journal = {ACM Computing Surveys},
    author = {Fagerholm, Fabian and Felderer, Michael and Fucci, Davide and Unterkalmsteiner, Michael and Marculescu, Bogdan and Martini, Markus and Tengberg, Lars Göran Wallgren and Feldt, Robert and Lehtelä, Bettina and Nagyváradi, Balázs and Khattak, Jehan},
    month = jan,
    year = {2022},
    pages = {1--36},
}

@article{russo2024NavigatingComplexityGenerative,
    title = {Navigating the {Complexity} of {Generative} {AI} {Adoption} in {Software} {Engineering}},
    volume = {33},
    issn = {1049-331X},
    url = {https://dl.acm.org/doi/10.1145/3652154},
    doi = {10.1145/3652154},
    number = {5},
    urldate = {2024-08-26},
    journal = {ACM Trans. Softw. Eng. Methodol.},
    author = {Russo, Daniel},
    month = jun,
    year = {2024},
    pages = {135:1--135:50},
}

@article{russo2022GenderDifferencesPersonality,
    title = {Gender {Differences} in {Personality} {Traits} of {Software} {Engineers}},
    volume = {48},
    issn = {1939-3520},
    url = {https://ieeexplore.ieee.org/document/9120355/?arnumber=9120355},
    doi = {10.1109/TSE.2020.3003413},
    number = {3},
    urldate = {2025-01-06},
    journal = {IEEE Transactions on Software Engineering},
    author = {Russo, Daniel and Stol, Klaas-Jan},
    month = mar,
    year = {2022},
    note = {Conference Name: IEEE Transactions on Software Engineering},
    pages = {819--834},
}

@inproceedings{kaptelinin2012AffordancesHCIMediated,
    address = {Austin Texas USA},
    title = {Affordances in {HCI}: toward a mediated action perspective},
    isbn = {978-1-4503-1015-4},
    shorttitle = {Affordances in {HCI}},
    url = {https://dl.acm.org/doi/10.1145/2207676.2208541},
    doi = {10.1145/2207676.2208541},
    language = {en},
    urldate = {2026-03-06},
    booktitle = {Proceedings of the {SIGCHI} {Conference} on {Human} {Factors} in {Computing} {Systems}},
    publisher = {ACM},
    author = {Kaptelinin, Victor and Nardi, Bonnie},
    month = may,
    year = {2012},
    pages = {967--976},
}

@inproceedings{amershi2019,
    address = {Glasgow Scotland Uk},
    title = {Guidelines for {Human}-{AI} {Interaction}},
    isbn = {978-1-4503-5970-2},
    url = {https://dl.acm.org/doi/10.1145/3290605.3300233},
    doi = {10.1145/3290605.3300233},
    language = {en},
    urldate = {2024-05-26},
    booktitle = {Proceedings of the 2019 {CHI} {Conference} on {Human} {Factors} in {Computing} {Systems}},
    publisher = {ACM},
    author = {Amershi, Saleema and Weld, Dan and Vorvoreanu, Mihaela and Fourney, Adam and Nushi, Besmira and Collisson, Penny and Suh, Jina and Iqbal, Shamsi and Bennett, Paul N. and Inkpen, Kori and Teevan, Jaime and Kikin-Gil, Ruth and Horvitz, Eric},
    month = may,
    year = {2019},
    pages = {1--13},
}

@misc{tie2024LLMsAreImperfect,
    title = {{LLMs} are {Imperfect}, {Then} {What}? {An} {Empirical} {Study} on {LLM} {Failures} in {Software} {Engineering}},
    shorttitle = {{LLMs} are {Imperfect}, {Then} {What}?},
    url = {http://arxiv.org/abs/2411.09916},
    doi = {10.48550/arXiv.2411.09916},
    urldate = {2024-11-22},
    publisher = {arXiv},
    author = {Tie, Jiessie and Yao, Bingsheng and Li, Tianshi and Ahmed, Syed Ishtiaque and Wang, Dakuo and Zhou, Shurui},
    month = nov,
    year = {2024},
    note = {arXiv:2411.09916},
}

@inproceedings{nguyen2024,
    address = {Honolulu HI USA},
    title = {How {Beginning} {Programmers} and {Code} {LLMs} ({Mis})read {Each} {Other}},
    isbn = {979-8-4007-0330-0},
    url = {https://dl.acm.org/doi/10.1145/3613904.3642706},
    doi = {10.1145/3613904.3642706},
    language = {en},
    urldate = {2024-08-08},
    booktitle = {Proceedings of the {CHI} {Conference} on {Human} {Factors} in {Computing} {Systems}},
    publisher = {ACM},
    author = {Nguyen, Sydney and Babe, Hannah McLean and Zi, Yangtian and Guha, Arjun and Anderson, Carolyn Jane and Feldman, Molly Q},
    month = may,
    year = {2024},
    pages = {1--26},
}

@inproceedings{xiao2024DevGPTStudyingDeveloperChatGPT,
    title = {{DevGPT}: {Studying} {Developer}-{ChatGPT} {Conversations}},
    issn = {2574-3864},
    shorttitle = {{DevGPT}},
    url = {https://ieeexplore.ieee.org/document/10555646/?arnumber=10555646},
    urldate = {2024-12-11},
    booktitle = {2024 {IEEE}/{ACM} 21st {International} {Conference} on {Mining} {Software} {Repositories} ({MSR})},
    author = {Xiao, Tao and Treude, Christoph and Hata, Hideaki and Matsumoto, Kenichi},
    month = apr,
    year = {2024},
    pages = {227--230},
}

@inproceedings{liang2024LargeScaleSurveyUsability,
    address = {New York, NY, USA},
    series = {{ICSE} '24},
    title = {A {Large}-{Scale} {Survey} on the {Usability} of {AI} {Programming} {Assistants}: {Successes} and {Challenges}},
    isbn = {979-8-4007-0217-4},
    shorttitle = {A {Large}-{Scale} {Survey} on the {Usability} of {AI} {Programming} {Assistants}},
    url = {https://dl.acm.org/doi/10.1145/3597503.3608128},
    doi = {10.1145/3597503.3608128},
    urldate = {2024-08-21},
    booktitle = {Proceedings of the {IEEE}/{ACM} 46th {International} {Conference} on {Software} {Engineering}},
    publisher = {Association for Computing Machinery},
    author = {Liang, Jenny T. and Yang, Chenyang and Myers, Brad A.},
    month = feb,
    year = {2024},
    pages = {1--13},
}

@inproceedings{ross2023ProgrammersAssistantConversational,
    address = {Sydney NSW Australia},
    title = {The {Programmer}’s {Assistant}: {Conversational} {Interaction} with a {Large} {Language} {Model} for {Software} {Development}},
    isbn = {979-8-4007-0106-1},
    shorttitle = {The {Programmer}’s {Assistant}},
    url = {https://dl.acm.org/doi/10.1145/3581641.3584037},
    doi = {10.1145/3581641.3584037},
    language = {en},
    urldate = {2024-02-14},
    booktitle = {Proceedings of the 28th {International} {Conference} on {Intelligent} {User} {Interfaces}},
    publisher = {ACM},
    author = {Ross, Steven I. and Martinez, Fernando and Houde, Stephanie and Muller, Michael and Weisz, Justin D.},
    month = mar,
    year = {2023},
    pages = {491--514},
}

@misc{austin2021ProgramSynthesisLarge,
    title = {Program {Synthesis} with {Large} {Language} {Models}},
    url = {http://arxiv.org/abs/2108.07732},
    language = {en},
    urldate = {2024-09-21},
    publisher = {arXiv},
    author = {Austin, Jacob and Odena, Augustus and Nye, Maxwell and Bosma, Maarten and Michalewski, Henryk and Dohan, David and Jiang, Ellen and Cai, Carrie and Terry, Michael and Le, Quoc and Sutton, Charles},
    month = aug,
    year = {2021},
    note = {arXiv:2108.07732 [cs]},
}

@article{hou2024LargeLanguageModels,
    title = {Large {Language} {Models} for {Software} {Engineering}: {A} {Systematic} {Literature} {Review}},
    issn = {1049-331X, 1557-7392},
    shorttitle = {Large {Language} {Models} for {Software} {Engineering}},
    url = {https://dl.acm.org/doi/10.1145/3695988},
    doi = {10.1145/3695988},
    language = {en},
    urldate = {2024-11-19},
    journal = {ACM Transactions on Software Engineering and Methodology},
    author = {Hou, Xinyi and Zhao, Yanjie and Liu, Yue and Yang, Zhou and Wang, Kailong and Li, Li and Luo, Xiapu and Lo, David and Grundy, John and Wang, Haoyu},
    month = sep,
    year = {2024},
    pages = {3695988},
}

@misc{draxler2023GenderAgeTechnology,
    title = {Gender, {Age}, and {Technology} {Education} {Influence} the {Adoption} and {Appropriation} of {LLMs}},
    url = {http://arxiv.org/abs/2310.06556},
    doi = {10.48550/arXiv.2310.06556},
    urldate = {2025-04-14},
    publisher = {arXiv},
    author = {Draxler, Fiona and Buschek, Daniel and Tavast, Mikke and Hämäläinen, Perttu and Schmidt, Albrecht and Kulshrestha, Juhi and Welsch, Robin},
    month = oct,
    year = {2023},
    note = {arXiv:2310.06556 [cs]},
}

@inproceedings{vasilescu2015GenderTenureDiversity,
    address = {Seoul Republic of Korea},
    title = {Gender and {Tenure} {Diversity} in {GitHub} {Teams}},
    isbn = {978-1-4503-3145-6},
    url = {https://dl.acm.org/doi/10.1145/2702123.2702549},
    doi = {10.1145/2702123.2702549},
    language = {en},
    urldate = {2024-09-19},
    booktitle = {Proceedings of the 33rd {Annual} {ACM} {Conference} on {Human} {Factors} in {Computing} {Systems}},
    publisher = {ACM},
    author = {Vasilescu, Bogdan and Posnett, Daryl and Ray, Baishakhi and Van Den Brand, Mark G.J. and Serebrenik, Alexander and Devanbu, Premkumar and Filkov, Vladimir},
    month = apr,
    year = {2015},
    pages = {3789--3798},
}

@inproceedings{pieterse2006SoftwareEngineeringTeam,
    address = {Somerset West South Africa},
    title = {Software engineering team diversity and performance},
    isbn = {978-1-59593-567-0},
    url = {https://dl.acm.org/doi/10.1145/1216262.1216282},
    doi = {10.1145/1216262.1216282},
    language = {en},
    urldate = {2026-03-06},
    booktitle = {Proceedings of the 2006 annual research conference of the {South} {African} institute of computer scientists and information technologists on {IT} research in developing countries},
    publisher = {South African Institute for Computer Scientists and Information Technologists},
    author = {Pieterse, Vreda and Kourie, Derrick G. and Sonnekus, Inge P.},
    month = oct,
    year = {2006},
    pages = {180--186},
}

@article{capretz2010WhyWeNeed,
    title = {Why do we need personality diversity in software engineering?},
    volume = {35},
    issn = {0163-5948},
    url = {https://dl.acm.org/doi/10.1145/1734103.1734111},
    doi = {10.1145/1734103.1734111},
    language = {en},
    number = {2},
    urldate = {2026-03-06},
    journal = {ACM SIGSOFT Software Engineering Notes},
    author = {Capretz, Luiz Fernando and Ahmed, Faheem},
    month = mar,
    year = {2010},
    pages = {1--11},
}

@article{davis1989,
    title = {Perceived {Usefulness}, {Perceived} {Ease} of {Use}, and {User} {Acceptance} of {Information} {Technology}},
    volume = {13},
    issn = {0276-7783},
    url = {https://www.jstor.org/stable/249008},
    doi = {10.2307/249008},
    number = {3},
    urldate = {2024-06-28},
    journal = {MIS Quarterly},
    publisher = {Management Information Systems Research Center, University of Minnesota},
    author = {Davis, Fred D.},
    year = {1989},
    pages = {319--340},
}

@incollection{hartDevelopmentNASATLXTask1988,
    series = {Human {Mental} {Workload}},
    title = {Development of {NASA}-{TLX} ({Task} {Load} {Index}): {Results} of {Empirical} and {Theoretical} {Research}},
    volume = {52},
    shorttitle = {Development of {NASA}-{TLX} ({Task} {Load} {Index})},
    url = {https://www.sciencedirect.com/science/article/pii/S0166411508623869},
    doi = {10.1016/S0166-4115(08)62386-9},
    urldate = {2024-06-28},
    booktitle = {Advances in {Psychology}},
    publisher = {North-Holland},
    author = {Hart, Sandra G. and Staveland, Lowell E.},
    editor = {Hancock, Peter A. and Meshkati, Najmedin},
    month = jan,
    year = {1988},
    pages = {139--183},
}

@article{vansomeren1994ThinkAloudMethod,
    title = {The think aloud method: a practical approach to modelling cognitive processes},
    volume = {11},
    shorttitle = {The think aloud method},
    url = {https://pure.uva.nl/ws/files/716505/149552_Think_aloud_method.pdf},
    number = {6},
    urldate = {2026-03-07},
    journal = {London: AcademicPress},
    author = {Van Someren, Maarten W. and Barnard, Yvonne F. and Sandberg, Jacobijn AC},
    year = {1994},
}

@book{corbin2015BasicsQualitativeResearch,
    address = {Los Angeles},
    edition = {Fourth},
    title = {Basics of qualitative research: techniques and procedures for developing grounded theory},
    isbn = {978-1-4129-9746-1},
    shorttitle = {Basics of qualitative research},
    language = {en},
    publisher = {SAGE},
    author = {Corbin, Juliet M. and Strauss, Anselm L.},
    year = {2015},
}

@article{khojah2024,
  title = {Beyond {{Code Generation}}: {{An Observational Study}} of {{ChatGPT Usage}} in {{Software Engineering Practice}}},
  shorttitle = {Beyond {{Code Generation}}},
  author = {Khojah, Ranim and Mohamad, Mazen and Leitner, Philipp and De Oliveira Neto, Francisco Gomes},
  year = 2024,
  month = jul,
  journal = {Proceedings of the ACM on Software Engineering},
  volume = {1},
  number = {FSE},
  pages = {1819--1840},
  issn = {2994-970X},
  doi = {10.1145/3660788},
  urldate = {2024-08-07},
  langid = {english}
}

@misc{shin2025PromptEngineeringFineTuning,
  title = {Prompt {{Engineering}} or {{Fine-Tuning}}: {{An Empirical Assessment}} of {{LLMs}} for {{Code}}},
  shorttitle = {Prompt {{Engineering}} or {{Fine-Tuning}}},
  author = {Shin, Jiho and Tang, Clark and Mohati, Tahmineh and Nayebi, Maleknaz and Wang, Song and Hemmati, Hadi},
  year = 2025,
  month = feb,
  number = {arXiv:2310.10508},
  eprint = {2310.10508},
  primaryclass = {cs.SE},
  publisher = {arXiv},
  doi = {10.48550/arXiv.2310.10508},
  urldate = {2026-06-16},
  archiveprefix = {arXiv},
  howpublished = {arXiv:2310.10508}
}

@incollection{hamid2024measure,
    title = {How to measure diversity actionably in technology},
    booktitle = {Equity, diversity, and inclusion in software engineering: {Best} practices and insights},
    publisher = {Apress Berkeley, CA},
    author = {Hamid, Md Montaser and Chatterjee, Amreeta and Guizani, Mariam and Anderson, Andrew and Moussaoui, Fatima and Yang, Sarah and Escobar, Isaac and Sarma, Anita and Burnett, Margaret},
    year = {2024},
    pages = {469--485},
}

@inproceedings{choudhuri2024HowFarAre,
    address = {Lisbon Portugal},
    title = {How {Far} {Are} {We}? {The} {Triumphs} and {Trials} of {Generative} {AI} in {Learning} {Software} {Engineering}},
    isbn = {979-8-4007-0217-4},
    shorttitle = {How {Far} {Are} {We}?},
    url = {https://dl.acm.org/doi/10.1145/3597503.3639201},
    doi = {10.1145/3597503.3639201},
    language = {en},
    urldate = {2024-09-21},
    booktitle = {Proceedings of the {IEEE}/{ACM} 46th {International} {Conference} on {Software} {Engineering}},
    publisher = {ACM},
    author = {Choudhuri, Rudrajit and Liu, Dylan and Steinmacher, Igor and Gerosa, Marco and Sarma, Anita},
    month = apr,
    year = {2024},
    pages = {1--13},
}

@article{pawlowsky-glahnLectureNotesCompositional,
    title = {Lecture {Notes} on {Compositional} {Data} {Analysis}},
    language = {en},
    author = {Pawlowsky-Glahn, Vera and Egozcue, Juan Jos and Tolosana-Delgado, Raimon},
}

@inproceedings{huynh2008DiscoveryActivityPatterns,
    address = {Seoul Korea},
    title = {Discovery of activity patterns using topic models},
    isbn = {978-1-60558-136-1},
    url = {https://dl.acm.org/doi/10.1145/1409635.1409638},
    doi = {10.1145/1409635.1409638},
    language = {en},
    urldate = {2026-03-07},
    booktitle = {Proceedings of the 10th international conference on {Ubiquitous} computing},
    publisher = {ACM},
    author = {Huynh, Tâm and Fritz, Mario and Schiele, Bernt},
    month = sep,
    year = {2008},
    pages = {10--19},
}

@inproceedings{kazemitabaar2023StudyingEffectAI,
    address = {Hamburg Germany},
    title = {Studying the effect of {AI} {Code} {Generators} on {Supporting} {Novice} {Learners} in {Introductory} {Programming}},
    isbn = {978-1-4503-9421-5},
    url = {https://dl.acm.org/doi/10.1145/3544548.3580919},
    doi = {10.1145/3544548.3580919},
    language = {en},
    urldate = {2026-03-07},
    booktitle = {Proceedings of the 2023 {CHI} {Conference} on {Human} {Factors} in {Computing} {Systems}},
    publisher = {ACM},
    author = {Kazemitabaar, Majeed and Chow, Justin and Ma, Carl Ka To and Ericson, Barbara J. and Weintrop, David and Grossman, Tovi},
    month = apr,
    year = {2023},
    pages = {1--23},
}
\end{document}